\documentclass[pra,showpacs,twocolumn]{revtex4}

\newcommand{\ket}[1]{|{#1}\rangle}
\newcommand{\bra}[1]{\langle{#1}|}

\newcommand{\be}{\begin{equation}}
\newcommand{\ee}{\end{equation}}

\usepackage{amsfonts}
\usepackage{amsmath}
\usepackage{graphicx}
\usepackage{amssymb}
\usepackage{amsmath}
\usepackage{amssymb}
\usepackage{graphicx}
\usepackage{lscape}
\usepackage{color}
\usepackage{epstopdf}

\begin{document}
\title{Single-branch theory of ultracold Fermi gases with artificial Rashba spin-orbit coupling}

\author{Daniel Maldonado-Mundo}
\affiliation{SUPA, Institute of Photonics and Quantum Sciences, Heriot-Watt University, Edinburgh EH14 4AS, United Kingdom}
\author{Patrik \"Ohberg}
\affiliation{SUPA, Institute of Photonics and Quantum Sciences, Heriot-Watt University, Edinburgh EH14 4AS, United Kingdom}
\author{Manuel Valiente}
\affiliation{SUPA, Institute of Photonics and Quantum Sciences, Heriot-Watt University, Edinburgh EH14 4AS, United Kingdom}
\begin{abstract}
We consider interacting ultracold fermions subject to Rashba spin-orbit coupling. We construct a single-branch interacting theory for the Fermi gas when the system is dilute enough so that the positive helicity branch is not occupied at all in the non-interacting ground state. We show that the theory is renormalizable in perturbation theory and therefore yields a model of polarized fermions that avoids a multi-channel treatment of the problem. Our results open the path towards a much more straightforward approach to the many-body physics of cold atoms subject to artificial vector potentials.
 
\end{abstract}
\maketitle
\section{Introduction}
The generation of artificial gauge potentials for ultracold atoms by means of properly engineered laser fields \cite{PatrikReview,Juzeliunas2004,Ruseckas2005,JakschZoller} constitutes a promising route towards the simulation of, for instance, Abelian \cite{Banerjee1} and non-Abelian \cite{Cirac,Zollernonabelian} gauge theories. Interestingly, before we are able to reach that major goal, many other interesting systems have been proposed, such as the simulation \cite{Matthew} of a bosonic interacting gauge theory \cite{Jackiw}, while some others have even been already implemented experimentally. These include the generation of a uniform Abelian vector potential \cite{Lin2009} leading to the observation of quantized vortices in a Bose-Einstein condensate \cite{Lin2009-2}, and spin-orbit coupling \cite{Lin2011}.  

Systems of ultracold spin-orbit coupled atoms have gained major interest recently. The phase diagram for weakly interacting bosons has been elucidated both in the untrapped case \cite{Zhai}, where the system exhibits plane-wave and striped phases, and in the case of a harmonically trapped Bose gas \cite{Santos}, where the phase diagram is even richer. In the case of fermions, the strength of the spin-orbit coupling plays a major role in the formation of two-body bound states \cite{Shenoy}, which exist even in the negative side of the s-wave scattering length in vacuum. Moreover, the crossover between a Bose-Einstein condensate (BEC) of molecules and a Bardeen-Cooper-Schrieffer (BCS) superfluid state can be driven by the spin-orbit coupling alone when the s-wave scattering length is fixed (for an extensive overview, see \cite{He}).

The many-body problems with and without spin-orbit coupling are very different from each other. For spin-$1/2$ fermions, the spin-orbit term couples the two $z$-components of the spin, turning two-particle scattering into a genuine multi-channel process \cite{ozawa}. For a dilute Fermi gas, the non-interacting lowest-energy channel is strongly preferred. However, as interactions are turned on, a single-channel description breaks down as a non-trivial, seemingly non-renormalizable ultraviolet (UV) behavior appears \cite{sachdev}. The full multi-channel description of the many-body problem is free of such anomalous UV structure, as shown by Ozawa and Baym in \cite{ozawa}. 

The multi-channel problem is a formidable one, hence, a renormalizable single-channel theory would greatly simplify the many-body problem. We here construct such a theory via perturbative renormalization of the two-body interaction, in a way that single-channel scattering reproduces the corresponding component of the exact T-matrix. We then obtain the energy of the Fermi gas in the normal phase to second order in the renormalized coupling constant, which is finite and independent of any momentum scales, thus showing that the theory is renormalizable.  
\section{Single-particle spectrum}
We begin by briefly reviewing the single-particle problem, which also sets the notation for the rest of the article. The most general single-particle problem with spin-orbit coupling can be diagonalized exactly \cite{Trushin}. We restrict ourselves, for concreteness, to a particular system of interest, with spin-orbit coupling of the Rashba type. The single-particle Hamiltonian is given by
\begin{equation}
H_0 =\frac{\mathbf{p}^2+\lambda^2}{2m}\hat{1}+\frac{\lambda}{m}\mathbf{\sigma}\cdot \mathbf{p}_{\perp},
\end{equation}
where $\mathbf{\sigma}=(\sigma_x,\sigma_y,\sigma_z)$ is the vector of spin-$1/2$ Pauli matrices. The helicity $\mathcal{H}\equiv \mathbf{\sigma}\cdot \mathbf{p}_{\perp}$, in the $x-y$ plane, is defined as the component of the spin in the direction of the in-plane momentum $\mathbf{p}_{\perp}=(p_x,p_y,0)$. The corresponding eigenvectors $\ket{\psi^{(\pm)}}$ define what we call the helicity basis. Since $[\mathcal{H},H_0]=0$, helicity is a good quantum number of the system, and we work, from now on, in the helicity basis. The two eigenvalues $h$ of the helicity for momentum $\mathbf{p}_{\perp}$ are given by $h_{\pm}=\pm p_{\perp}$, and the corresponding eigenstates are given by
\begin{equation}
\ket{\psi^{(\pm)}(\mathbf{r}_{\perp})}=e^{i(k_xx+k_yy)}\ket{\uparrow}\pm e^{i\gamma_k}e^{i(k_xx+k_yy)}\ket{\downarrow}.
\end{equation} 
where $\gamma_k$ is the polar angle of $\mathbf{p}_{\perp}$, given by $\tan{\gamma_k}=k_y/k_x$. The eigenstates of the single-particle Hamiltonian are therefore given by
\begin{equation}
\ket{\psi^{(\pm)}(\mathbf{r})}=\ket{\psi^{(\pm)}(\mathbf{r}_{\perp})}e^{ik_zz},
\end{equation}
and the energy dispersions $\epsilon_{\pm}$ have the form
\begin{equation}
\epsilon_{\pm}(\mathbf{p})=\frac{(p_{\perp}\pm \lambda)^2+p_z^2}{2m}.
\end{equation}.
It is also convenient to define spin states of the form
\begin{equation}
\ket{\pm,\hat{k}_{\perp}}=\ket{\pm}=\frac{\ket{\uparrow}\pm e^{i\gamma_k}\ket{\downarrow}}{\sqrt{2}},
\end{equation}
with $\hat{k}_{\perp}=\mathbf{k}_{\perp}/k_{\perp}$, so that $\ket{\psi^{(\pm)}_{\mathbf{k}}}=\ket{\mathbf{k}}\otimes\ket{\pm}=\ket{\mathbf{k},\pm}$. In the following, we will refer to negative (positive) helicity fermions as lower-branch (upper-branch) fermions.

\section{Free Fermi gas}
Before we proceed with the inclusion of interactions, we study the ground state of the free Fermi gas. We are aiming at constructing a single branch theory. Therefore, we must be in the regime where the fermions in the non-interacting ground state only occupy the  lower branch. Each fermion must therefore have an energy satisfying
\begin{equation}
\epsilon_{-}(\mathbf{p})\le E_F \le \frac{\lambda^2}{2m},
\end{equation}
where $E_F$ is the Fermi energy. We define, for convenience $E_F=k_F^2/2m$ (with $\hbar=1$ throughout), although this definition is arbitrary, since the single-particle dispersion is not quadratic. The above condition gives the region of integration for the $z$-component $q_z$ of the momentum in the energy states
\begin{equation}
|q_z|\le \sqrt{k_F^2-(q_{\perp}-\lambda)^2}.
\end{equation}
The integration interval for $q_{\perp}$ is given by $(\lambda-k_F,\lambda+k_F)$, which is {\it twice} as large as the Fermi momentum. The $q_z=0$ states therefore reach the energy $E_F$ when $q_{\perp}$ equals either end of the interval. The ground state energy is then given by
\begin{equation}
E_0 = \frac{2V}{(2\pi)^2}\int_{\lambda-k_F}^{\lambda+k_F}dq_{\perp}q_{\perp}\int_{0}^{\sqrt{k_F^2-(q_{\perp}-\lambda)^2}}dq_z \epsilon_{-}(\mathbf{q}).
\end{equation}
After a tedious but straightforward integration, we obtain
\begin{equation}
\frac{E_0}{V}=\frac{\lambda k_F^4}{16\pi m}=\frac{\pi}{m\lambda}\rho^2,
\end{equation}
where the density is given by
\begin{equation}
\rho=\frac{\lambda k_F^2}{4\pi}.
\end{equation}
It is interesting to note that the ground-state energy vanishes, for large $\lambda$, as $E_0\sim 1/\lambda$ in three-dimensions, while it would behave as $1/\lambda^2$ in two dimensions. This is due to the fact that in 3D finite-density states need to acquire momentum in the $z$-direction, and therefore there are less states with energies close to zero as compared to the 2D case where virtually every occupied single-particle state has vanishing energy.

\section{Effective interaction}
Several works on interacting bosons and fermions under spin-orbit coupling have noticed the appearance of an undesired ultraviolet (UV) logarithmic divergence in the T-matrix \cite{ozawa}, both in the vacuum and in the medium, when all other helicity channels, which include intraband interactions and interband interactions and transitions, are not taken into account. As shown by Ozawa and Baym \cite{ozawa}, the cancellation of the logarithmic divergence comes from the double transition (DT) process of two upper-branch fermions changing their helicity to $(-)$ ($\ket{++}\rightarrow\ket{--}$). Excluding the DT process in the effective theory means that the logarithmic divergence has to be explicitly eliminated. This can be done elegantly by introducing a pseudo-potential {\it a la} Fermi-Huang \cite{Huang} in momentum space \cite{Tanenergetics,ValienteTan}. All other collision processes, and the finite contribution of the DT process, are taken into account in the renormalization of the interaction.

\subsection{Logarithmic divergence}
The interaction between fermions in the lower-branch is given by \cite{ozawa}
\begin{equation}
\hat{V}=\frac{g_*}{2V}\sum_{\mathbf{p}_1+\mathbf{p}_2=\mathbf{p}_3+\mathbf{p}_4}\Lambda(\mathbf{p})\Delta(\gamma_1,\gamma_2,\gamma_3,\gamma_4) c_{\mathbf{p}_4}^{\dagger}c_{\mathbf{p}_3}^{\dagger}c_{\mathbf{p}_2}c_{\mathbf{p}_1},\label{bareinteraction}
\end{equation}
where $c_{\mathbf{p}}$ annihilates a fermion in state $\ket{\mathbf{p},-}$, $g_*$ is the bare (un-renormalized) coupling constant, and 
\begin{equation}
\Delta(\gamma_1,\gamma_2,\gamma_3,\gamma_4) = -\frac{1}{8}(e^{i\gamma_1}-e^{i\gamma_2})(e^{-i\gamma_3}-e^{-i\gamma_4}).
\end{equation}
The relative momentum $\mathbf{p}$ is defined by $\mathbf{p}=(\mathbf{p}_2-\mathbf{p}_1)/2$, with $\gamma_i\equiv \gamma_{\mathbf{p}_i}$, while $\Lambda(\mathbf{p})$ is Tan's $\Lambda$-distribution \cite{Tanenergetics}, given by \cite{ValienteTan}
\begin{equation}
\Lambda(\mathbf{p})=1-\frac{\delta(1/p)}{p}.
\end{equation}
Note that by including Tan's distribution in the interaction in Eq. (\ref{bareinteraction}), the linear ultraviolet (UV) divergence connected to the s-wave scattering length is already absent. The logarithmic divergence in the single-branch model comes from attempting to calculate the two-body T-matrix using the single-channel approach. To see this, we notice that the single-channel second Born approximation is given by
\begin{align}
&T^*(\mathbf{p},\mathbf{p'};\mathbf{Q})=g_*\Delta(\gamma_1,\gamma_2,\gamma_3,\gamma_4)\nonumber \\
&+ g_*^2\int \frac{d\mathbf{k}}{(2\pi)^3}\Lambda(\mathbf{k})\frac{\Delta(\gamma_1,\gamma_2,\gamma_5,\gamma_6)\Delta(\gamma_5,\gamma_6,\gamma_3,\gamma_4)}{\epsilon_{-}(\frac{\mathbf{q}}{2}-\mathbf{k})+\epsilon_{-}(\frac{\mathbf{q}}{2}+\mathbf{k})}+O(g_*^3). \label{TmatrixBorn}
\end{align}
In the above equation, $\mathbf{Q}$ is the center-of-mass momentum, and $\mathbf{p}$ and $\mathbf{p}'$ are, respectively, the incoming and outgoing relative momenta of the two-body system. The angles $\gamma_1$, $\gamma_2$, $\gamma_3$, $\gamma_4$, $\gamma_5$ and $\gamma_6$ correspond, respectively to the momenta $\mathbf{Q}/2+\mathbf{p}$, $\mathbf{Q}/2-\mathbf{p}$, $\mathbf{Q}/2-\mathbf{p}'$, $\mathbf{Q}/2+\mathbf{p}'$, $\mathbf{Q}/2-\mathbf{k}$ and $\mathbf{Q}/2+\mathbf{k}$ . As it stands, Eq. (\ref{TmatrixBorn}) is free of the linear UV divergence (which is removed by the pseudopotential), but has a logarithmic divergence. After trivial algebraic manipulations, we obtain from Eq. (\ref{TmatrixBorn})
\begin{align}
&T^*(\mathbf{p},\mathbf{p'};\mathbf{Q})=g_*\Delta(\gamma_1,\gamma_2,\gamma_3,\gamma_4) \times \nonumber \\
&\left(1 - \frac{g_*}{8}\int \frac{d\mathbf{k}}{(2\pi)^3}\Lambda(\mathbf{k})\frac{2[1-\cos(\gamma_5-\gamma_6)]}{\epsilon_{-}(\frac{\mathbf{Q}}{2}-\mathbf{k})+\epsilon_{-}(\frac{\mathbf{Q}}{2}+\mathbf{k})}\right)+O(g_*^3).\label{Tstar}
\end{align} 
The UV structure of the T-matrix at a large momentum cutoff $\beta$, is given by
\begin{equation}
\frac{T^*}{g_*\Delta}\sim 1 - \frac{mg_*}{2\pi\hbar^2}\lambda\log\frac{\beta}{\beta_0}+\ldots 
\end{equation}  
where $\beta_0$ is a finite, arbitrary momentum scale with the only purpose of rendering the argument of the logarithm dimensionless. The logarithmic divergence above is the one we aim at eliminating in a physically consistent way, that is, renormalizing it away without the appearance of any extra scales or fitting parameters in the system. We will show in the next section that this is indeed possible at the many-body level.

\subsection{Renormalization of the coupling constant}
The single-channel T-matrix, $T^*$, calculated in the previous subsection is still not correct at second order in the coupling constant. We therefore invoke perturbative renormalization so that the single-channel model reproduces the correct T-matrix. This is given by \cite{ozawa}
\begin{equation}  
T(\mathbf{p},\mathbf{p}';\mathbf{Q})=\frac{g\Delta(\gamma_1,\gamma_2,\gamma_3,\gamma_4)}{1+\frac{mg\lambda}{4\pi}[A(\tilde{Q}/2)+B(\tilde{Q}/2)]},\label{exactTmatrix}
\end{equation}
where $g=4\pi a/m$ is the renormalized coupling constant, with $a$ the s-wave scattering length, $\tilde{Q}=Q/\lambda$ and $A$ and $B$ are given by
\begin{align}
A(\tilde{Q}/2)&=\frac{\pi}{m\lambda}\int\frac{d\mathbf{k}}{(2\pi)^3}\Lambda(\mathbf{k})\left[\frac{1}{\epsilon_{-}\left(\frac{\mathbf{Q}}{2}+\mathbf{k}\right)+\epsilon_{-}\left(\frac{\mathbf{Q}}{2}-\mathbf{k}\right)}\right.\nonumber \\
&+\frac{1}{\epsilon_{+}\left(\frac{\mathbf{Q}}{2}+\mathbf{k}\right)+\epsilon_{+}\left(\frac{\mathbf{Q}}{2}-\mathbf{k}\right)}\nonumber \\
&\left. +\frac{2}{\epsilon_{-}\left(\frac{\mathbf{Q}}{2}+\mathbf{k}\right)+\epsilon_{+}\left(\frac{\mathbf{Q}}{2}-\mathbf{k}\right)}\right],
\end{align}
and
\begin{align}
B(\tilde{Q}/2)&=-\frac{\pi}{m\lambda}\int\frac{d\mathbf{k}}{(2\pi)^3}\Lambda(\mathbf{k})\cos(\gamma_5-\gamma_6)\times \nonumber \\
&\left[\frac{1}{\epsilon_{-}\left(\frac{\mathbf{Q}}{2}+\mathbf{k}\right)+\epsilon_{-}\left(\frac{\mathbf{Q}}{2}-\mathbf{k}\right)}\right.\nonumber \\
&+\frac{1}{\epsilon_{+}\left(\frac{\mathbf{Q}}{2}+\mathbf{k}\right)+\epsilon_{+}\left(\frac{\mathbf{Q}}{2}-\mathbf{k}\right)}\nonumber \\
&\left. -\frac{2}{\epsilon_{-}\left(\frac{\mathbf{Q}}{2}+\mathbf{k}\right)+\epsilon_{+}\left(\frac{\mathbf{Q}}{2}-\mathbf{k}\right)}\right]
\end{align}
The single-channel T-matrix in Eq. (\ref{Tstar}) is obviously correct to first order in $g$. However, the second-order term is not correct and logarithmically divergent, and needs to be renormalized. In fact, the logarithmic divergence can be safely eliminated at this stage using a pseudopotential, as we will show in the last section of the article, but we first show that it is cancelled in the course of renormalization. 
We begin the renormalization process by expanding the bare coupling constant in powers of its renormalized counterpart
\begin{equation}
g_*(\lambda,Q)=g+\alpha_{\lambda,Q}g^2+O(g^3).\label{renorm}
\end{equation}
The renormalization condition reads 
\begin{equation}
T(\mathbf{p},\mathbf{p}';\mathbf{Q})=T^*(\mathbf{p},\mathbf{p}';\mathbf{Q}),
\end{equation}
to the given order in $g$. Expanding the exact T-matrix, Eq. (\ref{exactTmatrix}), to second order, and equating it to the single-channel T-matrix, Eq. (\ref{Tstar}), we obtain 
\begin{equation}
\alpha_{\lambda,Q}=\frac{\mathcal{I}_{\lambda,Q}}{8}-\frac{m\lambda}{4\pi\hbar^{\alpha}}[A(\tilde{Q}/2)+B(\tilde{Q}/2)],
\end{equation}
where we have defined
\begin{equation}
\mathcal{I}_{\lambda,Q}=2\int\frac{d\mathbf{k}}{(2\pi)^3}\Lambda(\mathbf{k})\frac{1-\cos(\gamma_5-\gamma_6)}{\epsilon_{-}(\mathbf{Q}/2-\mathbf{k})+\epsilon_{-}(\mathbf{Q}/2+\mathbf{k})}.
\end{equation}
As can be readily checked, $\mathcal{I}_{\lambda,Q}$, and therefore $\alpha_{\lambda,Q}$, is logarithmically divergent.

\section{Interacting Fermi gas}
We develop here a perturbation theory for the interacting Fermi gas, using our perturbatively-renormalized single-channel theory, and show that the theory is indeed renormalizable and free of any logarithmic UV-divergence after renormalization. 

The first order contribution of the interaction is given by
\begin{align}
\frac{E^{(1)}}{V}&\equiv \bra{F}\hat{V}\ket{F} \nonumber \\
&= -\frac{g}{4}\frac{1}{(2\pi)^6}\int_{F}d\mathbf{q}\int_{F}d\mathbf{q}' \left[\cos(\gamma_{\mathbf{q}}-\gamma_{\mathbf{q}'})-1\right],
\end{align}
where $\ket{F}$ is the non-interacting ground state (Fermi sea) and the momenta in the integration are restricted to values within the Fermi sea, together with $g_*=g+O(g^2)$. The final result is
\begin{equation}
\frac{E^{(1)}}{V}=\frac{g}{4}\rho^2.
\end{equation}

At second order there are two different contributions. The first one is due to the renormalization process at order $g^2$ in Eq. (\ref{renorm}),
\begin{equation}
\frac{E^{(2)}_1}{V}=\left(\frac{g^2}{4}\right)\frac{1}{(2\pi)^6}\int_F d\mathbf{q}\int_Fd\mathbf{q}' [\cos(\gamma_{\mathbf{q}}-\gamma_{\mathbf{q}'})-1]\alpha_{\lambda,Q},
\end{equation}
where $\mathbf{Q}=\mathbf{q}+\mathbf{q}'$. The other contribution arises from the usual second-order diagrams and is given by
\begin{equation}
\frac{E^{(2)}_2}{V}=-\frac{g^2}{4(2\pi)^9}\int_Fd\mathbf{q}\int_Fd\mathbf{q}'\int_{\mathbf{k}\not\in F} d\mathbf{k}\frac{\Lambda(\mathbf{k})\mathcal{C}_{\mathbf{k},\mathbf{k}'}\mathcal{C}_{\mathbf{q},\mathbf{q}'}}{\epsilon_{-}(\mathbf{k},\mathbf{k}',\mathbf{q},\mathbf{q}')},\label{E22}
\end{equation} 
where $\mathbf{k}'=\mathbf{q}+\mathbf{q}'-\mathbf{k}$, $\mathcal{C}_{\mathbf{k},\mathbf{k}'}=1-\cos(\gamma_{\mathbf{k}}-\gamma_{\mathbf{k}'})$ and 
\begin{equation}
\epsilon_{-}(\mathbf{k},\mathbf{k}',\mathbf{q},\mathbf{q}')\equiv \epsilon_{-}(\mathbf{k})+\epsilon_{-}(\mathbf{k'})-\epsilon_{-}(\mathbf{q})-
\epsilon_{-}(\mathbf{q}').
\end{equation}

There is still one important point yet to prove, and that is the independence of the results on the renormalization prescription, so that there is no arbitrary scale left. The theory will then be free of UV-divergences and therefore has predictive power. We show here that the second order correction to the energy $E^{(2)}=E^{(2)}_1+E^{(2)}_2$ is finite by showing that the logarithmic divergences present in $E^{(2)}_1$ and $E^{(2)}_2$ cancel each other. To see this, notice that for large $\mathbf{k}$, we have $1/\epsilon_{-}(\mathbf{k},\mathbf{k}',\mathbf{q},\mathbf{q}')=1/2\epsilon_{-}(\mathbf{k})+O[(\epsilon_{-}(\mathbf{k}))^{-2}]$. All the divergent UV structure of the integral in Eq. (\ref{E22}) appears at this order alone. A formal expansion for $E^{(2)}_2$ then has the form
\begin{equation}
\frac{E^{(2)}_2}{V}=-\frac{g}{8(2\pi)^3}\rho^2\int_{\mathbf{k}\not\in F}d\mathbf{k}\frac{\Lambda(\mathbf{k})}{\epsilon_{-}(\mathbf{k})}+\ldots,
\end{equation}
while the divergent part of $E^{(2)}_{1}$ is easily isolated from the finite part. This is given by
\begin{equation}
\frac{E^{(2)}_1}{V}=\frac{g^2}{8(2\pi)^3}\rho^2\int_{\mathbf{k}\not\in F}d\mathbf{k}\frac{\Lambda(\mathbf{k})}{\epsilon_{-}(\mathbf{k})}+\ldots
\end{equation}
From the two equations above, we see that $E^{(2)}=E^{(2)}_1+E^{(2)}_2$ is not logarithmically divergent, which is what we wanted to show. The single-channel theory can therefore be nicely renormalized in favor of only the scattering length.

\section{Completely divergence-free approach}
All the results of the previous subsection can actually be reproduced with a theory that is free of the logarithmic divergence and is finite at all steps. To do so, we need to replace the pseudopotential $\Lambda(\mathbf{k})$ by a new pseudopotential $\tilde{\Lambda}(\mathbf{k})$, together with an irrelevant, arbitrary momentum scale $\tilde{\beta}_0$. The modified pseudopotential has the form
\begin{equation}
\tilde{\Lambda}(\mathbf{k})= \Lambda(\mathbf{k})-\frac{\lambda}{k_{\perp}}\log\left(\frac{k_{\perp}}{\tilde{\beta}_0}\right)\delta^{(2)}(1/k_{\perp}),
\end{equation}
where $\delta^{(2)}$ is the two-dimensional Dirac delta. It is easy to see that in this case $E^{(2)}_1$ and $E^{(2)}_2$ are both finite separately, and so is the bare coupling constant $g_*$, but both $E^{(2)}_1$ and $E^{(2)}_2$ depend on the choice of $\tilde{\beta}_0$. The renormalizability of the theory depends now on proving that the results are independent of the scale $\tilde{\beta}_0$, that is, to show that it is irrelevant. Proceeding in a way completely analogous to that of the previous section, we find that this is indeed the case.

\section{Conclusions and outlook}
We have introduced a renormalizable theory for interacting fermions subject to spin-orbit coupling. Our theory is valid in the dilute regime where the fermions in the non-interacting ground state occupy only the lower helicity branch. The effective single-branch model corresponds to interacting polarized fermions, and thus opens the path to a simpler treatment of the many-body problem, circumventing the intricacies of the full multi-channel system. We have illustrated our methods by calculating the second-order correction to the ground-state energy of the repulsive Fermi gas. 

As a natural extension of our work, it would be interesting to calculate the non-Hermitian optical potential, which takes explicitly into account the population of the other channels. Our model may also be of interest for a more sofisticated treatment of the BEC-BCS crossover, and for reduced dimensional Fermi and Bose gases subject to spin-orbit coupling.  

\acknowledgements{Useful correspondence with T. Ozawa is gratefully acknoledged. D.M.-M. acknowledges support from the EPSRC CM-DTC, P.\"O and M.V.  acknowledge support from EPSRC grant No. EP/J001392/1.}

\bibliographystyle{unsrt}

\end{document}